\documentclass[aps,prl,twocolumn]{revtex4}

\usepackage{graphicx,latexsym,color,amsmath}

\def\be{\begin{equation}}
\def\ee{\end{equation}}
\def\bea{\begin{eqnarray}}
\def\eea{\end{eqnarray}}
\def\bi{\begin{itemize}}
\def\ei{\end{itemize}}
\def\K{$^{40}$K}
\def\Rb{$^{87}$Rb}
\def\exp#1{\times 10^{#1}}
\def\ket#1{\left|#1\right\rangle}

\begin{document}
\hbadness = 10000

\title{Radio frequency association of heteronuclear Feshbach molecules}

\author{C. Klempt, T. Henninger, O. Topic, M. Scherer, L. Kattner, E. Tiemann, W. Ertmer and J. J. Arlt}
\affiliation{Institut f\"ur Quantenoptik, Leibniz Universit\"at Hannover,
Welfengarten 1, D-30167 Hannover, Germany}

\date{\today}

\begin{abstract}
We present a detailed analysis of the production efficiency of weakly bound heteronuclear $^{40}$K~$^{87}$Rb Feshbach molecules using radio frequency association in a harmonic trap. The efficiency was measured in a wide range of temperatures, binding energies and radio frequencies. A comprehensive analytical model is presented, explaining the observed asymmetric spectra and achieving good quantitative agreement with the measured production rates. This model provides a deep understanding of the molecule association process and paves the way for future experiments which rely on Feshbach molecules e.g. for the production of deeply bound molecules. 
\end{abstract}

\maketitle
The creation and investigation of ultracold, trapped, deeply bound molecules has been a major research goal for a number of years. In particular polar molecules with long range dipolar interaction  \cite{Santos2000} offer a number of fascinating experimental avenues such as the
investigation of quantum phases \cite{Goral2002}, methods for quantum computation \cite{DeMille2002} or precision measurements of fundamental constants \cite{Hudson2006}. Recently, a number of approaches have made significant progress towards that goal. Polar molecules have been slowed down and trapped using Stark deceleration in spatially modulated electric fields \cite{Crompvoets2001,Sawyer2007,Bucicov2008}. Other approaches are provided by cooling ground state molecules with a buffer gas \cite{Weinstein1998} and by photoassociation in magneto-optical traps, where the lowest vibrational level in ultracold $K_2$ \cite{Nikolov2000}, $KRb$ \cite{Mancini2004}, $RbCs$ \cite{Sage2005}, $Cs_2$ \cite{Viteau2008}, and $LiCs$ \cite{Deiglmayr2008} has been reached. However, these approaches only offer limited state control or temperatures in the mK regime. On the other hand, the understanding and control of interactions in ultracold gases has made dramatic progress in the past few years. The cooling procedure for two atomic species as well as the heteronuclear interaction potentials have been studied in detail \cite{Stan2004,Simoni2008,Taglieber2008,Wille2008}, especially in the case of \K- and \Rb-mixtures \cite{Inouye2004,Ferlaino2006,Ospelkaus2006a,Klempt2007}. This development now allows for the association of weakly bound heteronuclear molecules in well defined quantum states using magnetically tunable Feshbach resonances \cite{Ospelkaus2006,Papp2006,Zirbel2008}. A detailed understanding of the association process of these molecules is of central importance. These Feshbach molecules provide an ideal starting point for a further transfer into deeply bound states \cite{Winkler2007,Ospelkaus2008,Danzl2008}. Recently, the de-excitation of these molecules to the rovibrational ground state has also been demonstrated \cite{Ni2008}.

So far, the association of homonuclear molecules has been described with a semi-classical numerical model \cite{Hodby2005}, which does not predict absolute conversion rates. The model assumes adiabatic conversion of selected atom pairs to molecules. It has also been applied to heteronuclear mixtures \cite{Papp2006,Zirbel2008}. However, in this case the approach is only valid if the conversion speed is small enough to assure adiabaticity, while it has to be large enough such that the atom pairs do not drift apart. Further theoretical studies focus on the use of modulated magnetic fields for the production of molecules \cite{Hanna2007,Bertelsen2007}.

\begin{figure}[ht!]

\centering

\includegraphics*[width=8.6cm]{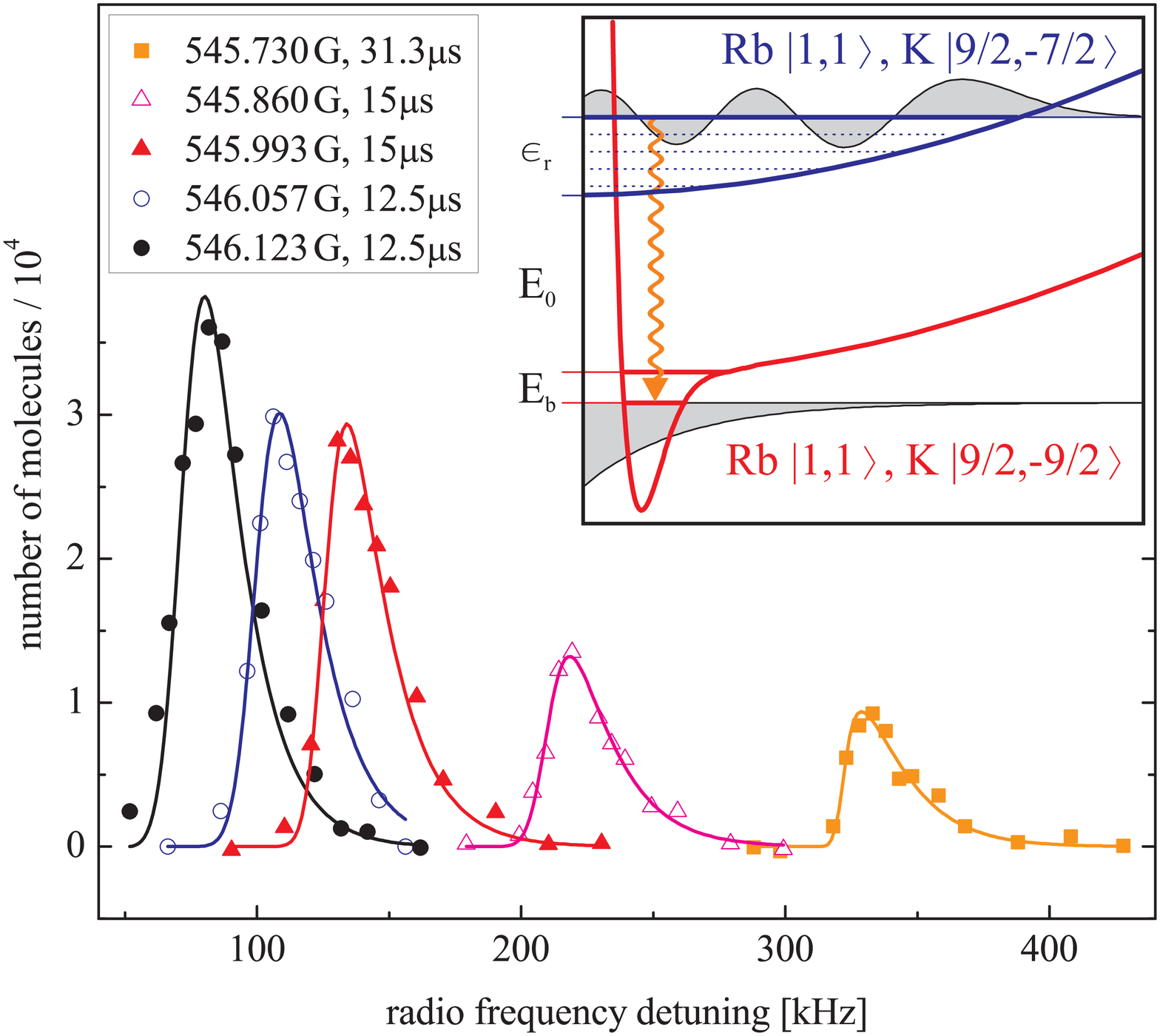}

\caption{Number of associated molecules versus radio frequency for various binding energies. The solid lines show the theoretical model. All measurements were taken at a temperature of $730\ nK$. The duration of the radio frequency pulses is indicated in the legend. The inset illustrates the association process including the relevant energies and the wave functions of molecule and atom pair.}

\label{fig1}

\end{figure}

In this paper, we present a complementary analytical model for radio frequency association which yields absolute molecular production rates. The model is compared to radio frequency association spectra obtained in an ultracold mixture of \Rb~ and \K~ atoms. It explains the measured asymmetric molecular association spectra depending on the energy distribution of the atoms in the trap and thus provides a deep understanding of the molecule association process. 
 
In our approach, the number of molecules is calculated by time dependent perturbation theory assuming a Gaussian time dependence of the coupling term. The treatment is analogous to the derivation of Fermi's Golden Rule, which has been successfully used to describe radio frequency dissociation \cite{Chin2004,Chin2005}. In the case of radio frequency association, the number of colliding pairs per energy interval $h(\epsilon_r)$ has to be included, where $\epsilon_r$ is the energy of the relative motion. In the limit of small depletion, the perturbative approach for short radio frequency pulses with a Gaussian shaped amplitude of duration $\tau$ yields the number of molecules.
\be
N_{mol}=\frac{\pi}{2} \Omega^2 \tau^2 \int_0^\infty h(\epsilon_r) e^{-\frac{(E_{rf}-E_b-E_0-\epsilon_r)^2 \tau^2}{\hbar^2}} F_f(\epsilon_r) d\epsilon_r
\label{eq:num_mol}
\ee

Here, $E_{rf}=h \nu$ is the energy corresponding to the radio frequency $\nu$ and $\Omega$ is the Rabi frequency of the atomic transition with energy $E_0$ (see inset of Fig. \ref{fig1}). $E_b$ denotes the molecular binding energy and

\be
F_f(\epsilon_r)= \left| \int \Psi^*_{\epsilon_r} (r) \Phi_m (r) dr \right|^2
\ee
is the Franck-Condon factor between the wave function $\Psi_{\epsilon_r} (r)$ of the colliding atom pair and the bound molecular wave function $\Phi_m (r)$. 

To evaluate equation \eqref{eq:num_mol}, we start with the occupation density $h(\epsilon_r)$ by considering the distributions of the two atomic species in the harmonic trap. The Hamiltonian of two noninteracting atoms a and b describes two uncoupled three-dimensional harmonic oscillators

\be
H_1=\frac{p_a{}^2}{2 m_a}+\frac{m_a}{2} \sum_i\omega _{a,i}{}^2x_{a,i}{}^2+\frac{p_b{}^2}{2 m_b}+\frac{m_b}{2} \sum_i\omega _{b,i}{}^2x_{b,i}{}^2,
\ee

where $p_{a/b}$ denotes the momenta of the atoms, $x_{a/b,i}$ the $i$th component of their positions, $m_{a/b}$ their masses and $\frac{1}{2 \pi} \omega_{a/b,i}$ the $i$th component of their trapping frequencies. One can introduce center-of-mass coordinates as usually done in free space, where $r, p, \mu$ describe position, momentum and mass of the relative motion and $R, P$ and $M$ the motion of the center of mass \cite{Bertelsen2007}.

The Hamiltonian then reads

\be
H_2=\frac{P^2}{2 M}+\frac{M}{2} \sum_i\bar{\omega}_i^2 R_i^2 +\frac{p^2}{2 \mu}+ \frac{\mu}{2} \sum_i \bar{\omega}_i^2 r_i^2 + H_{c},\\
\ee

where $\bar{\omega}_i=(\omega_{a,i} \omega_{b,i})^{1/2}$ is the mean trapping frequency. This Hamiltonian again contains two harmonic oscillators, one for the relative and one for the collective motion. The residual term $H_c$ couples relative and collective motion and is proportional to $(\omega_{a,i}-\omega_{b,i})$. If the trapping frequencies are similar, this coupling is small and will be neglected for the molecule production on short timescales.

The number of atoms per state and energy $\epsilon$ in a single-atom harmonic oscillator is assumed to be Maxwell-Boltzmann distributed in the thermodynamic limit $f(\epsilon)= N\frac{\hbar^3 \omega_1\omega_2\omega_3}{(k_B T)^3} exp(-\epsilon/(k_B T))$. Therefore, the number of possible atom pairs per state and total energy $\epsilon_t=\epsilon_a+\epsilon_b$ is simply the product of the two single occupation densities $f_{p}(\epsilon_t)=N_a N_b(\frac{\hbar \tilde{\omega}}{k_B T})^6 exp(-\epsilon_t/(k_B T))$, where $\tilde{\omega}=(\prod_i \bar{\omega}_i)^{1/3}$. Since the two Hamiltonians $H_1$ and $H_2$ describe the same system, degenerate states in the two-atoms-representation have to map the corresponding degenerate eigenstates in the atom-pair-representation homogeneously. Thus, the number of pairs per state $f_{p}(\epsilon_t)$ has to be equal for the  two representations. In order to obtain the total number per energy, the density of states has to be included, which is the product of the density of states of the relative and the collective motion. The three-dimensional harmonic oscillator of the center-of-mass motion has a density of states $g_{cm}(\epsilon_{cm})=\frac{\epsilon_{cm}^2}{2(\hbar \tilde{\omega})^3}$. Only atom pairs colliding with vanishing angular momentum contribute to the formation of s-wave Feshbach molecules. Therefore, the density of s-wave states per relative kinetic energy is given by $g_r(\epsilon_r)=\frac{1}{2 \hbar \tilde{\omega}}$. 

For the molecule production, only the energy of the relative motion $\epsilon_r$ is relevant, since the molecule follows the same center-of-mass motion as the atom pair. The number of molecules per energy in the relative motion can be calculated by integrating over all possible center-of-mass energy states.

\bea
h(\epsilon_r)&=&\int_{0}^\infty g_{cm}(\epsilon_{cm}) g_r(\epsilon_r) f_{p}(\epsilon_r+\epsilon_{cm}) d\epsilon_{cm} \nonumber \\
&=& N_a N_b \frac{(\hbar \tilde{\omega})^2}{2(k_B T)^3} e^{-\frac{\epsilon_r}{k_B T}}
\eea

For the complete evaluation of eq. \eqref{eq:num_mol}, we now evaluate the Franck-Condon factor $F_f(\epsilon_r)$. In the case of small interatomic distance, the wave function of the atom pair $\Psi^*_{\epsilon_r}$ matches the wave function of a free atom pair except for a factor of $\hbar \tilde{\omega}$ for correct normalization. To calculate the Franck-Condon factor, it is crucial to include the scattering length of the colliding atom pair $a'$, since the molecule production probes the innermost part of the scattering wave function. Nevertheless, it is valid to neglect the influence of the scattering length on the energy distribution at the relevant densities. Hence, we adopt the Franck-Condon factor from Ref.~\cite{Chin2005}. 

\be
F_f(\epsilon_r)= \hbar \tilde{\omega} \: \chi(E_b) \frac{2}{\pi} \left( 1 - \frac{a'}{a} \right)^2 \frac{\sqrt{\epsilon_r}\sqrt{E_b} E_b'}{(\epsilon_r+E_b)^2 (\epsilon_r+E_b')}
\ee

where $E_b$ is the binding energy of the formed molecule and according to Ref. \cite{Chin2005}, $a$ and $E_b'$ are defined implicitly by

\be
E_b \approx \frac{\hbar^2}{2 \mu a^2} \quad\quad\quad E_b' \approx \frac{\hbar^2}{2 \mu a'^2}.
\label{eq:B_of_a}
\ee

The factor $\chi(E_b)$ reflects that the molecular wave function consists of two parts, the closed channel and the open channel fraction. Only the latter contributes significantly to the Franck-Condon factor. It can be approximated \cite{Zirbel2008} by $\chi(E_b)=1-(1/\Delta\mu) \: \partial E_b/\partial B$, where $B$ denotes the magnetic field and $\Delta\mu$ is the difference in magnetic moment of the molecular and the atomic state. In the vicinity of the Feshbach resonance ($\rm{sign}(a_{bg})\frac{\Delta B}{B-B_0}>>1$), $E_b$ can be approximated using \eqref{eq:B_of_a} and the scattering length

\be
a(B)=a_{bg}(1-\Delta B/(B-B_0)).
\label{eq:scatteringlength}
\ee

Here, $a_{bg}$ denotes the background scattering length, $B_0$ and $\Delta B$ are the position and the width of the resonance. In total, one obtains
\be
\chi(E_b)=1-\frac{\hbar^2 k^2}{\Delta\mu \Delta B\: \mu \:a_{bg}}(1+k a_{bg})^2,
\label{eq:chi}
\ee
where $k^2=E_b/2 \mu$.

Now, all parts of equation \eqref{eq:num_mol} are derived. In the following, we directly fit the theoretical prediction to our experimental data, allowing only two free parameters: The binding energy $E_b$ and a scale factor $\lambda$ to quantify deviations from our model. For the atom pair scattering length, the values $a_{bg}=9.88\ nm$, $a'=9.10\ nm$ and $\Delta \mu = 2.32\ \mu_B$ are adopted from a coupled channel calculation \cite{Klempt2007,Pashov2007}.

We used the following experimental sequence to produce heteronuclear \K-\Rb-molecules close to a Feshbach resonance located at $547\ G$ and to measure their production efficiency. First, a mixture of $2\exp{6}$ \K-Atoms and $1\exp{6}$ \Rb-atoms at a temperature of $2\ \mu K$ is prepared in a crossed beam optical dipole trap at a wavelength of $1064\ nm$. We employ rapid adiabatic passages to transfer the \Rb~atoms to the $\ket{1,1}$ state and the \K~ atoms to the $\ket{9/2, -7/2}$ state at homogeneous magnetic fields of $10\ G$ and $20\ G$. Then the magnetic field is increased to values below the Fesh\-bach resonance. The sum of magnetic field noise, field inhomogeneity and shot-to-shot fluctuations is less than $16\ mG$, accomplished by a mechanical transport to the geometric center of the Feshbach coils \cite{Klempt2008}. The power of the dipole trap is lowered to mean trapping frequencies of $244~{\rm Hz}$ ($335~{\rm Hz}$) for \Rb~ (\K), yielding samples of $5\exp{5}$ \K~ atoms and $2.5\exp{5}$ \Rb~ atoms at a temperature of $730\ nK$. Finally, we apply a Gaussian shaped radio frequency pulse with a width of $\tau=12.5$ to $31.25\ \mu s$ and a peak atomic Rabi frequency of $\Omega=2 \pi\ 45\ kHz$ to produce up to $5\exp{4}$ weakly bound molecules (see inset of Fig. \ref{fig1}). The duration of the radio frequency pulse is chosen short enough to ensure that the production efficiency is still in the linear regime, i.e. the loss of molecules due to interaction with residual atoms can be neglected \cite{Zirbel2008a}.

Immediately after the molecule association, the trapping potential is switched off and the number of associated molecules is measured directly using resonant light for \K~ atoms in the state $\ket{9/2, -9/2}$ after 2 ms of ballistic expansion. In two subsequent absorption images, we measure the remaining number of \K~ and \Rb~ atoms in the states $\ket{9/2, -7/2}$ and $\ket{1,1}$.

\begin{figure}[ht!]

\centering

\includegraphics*[width=8.6cm]{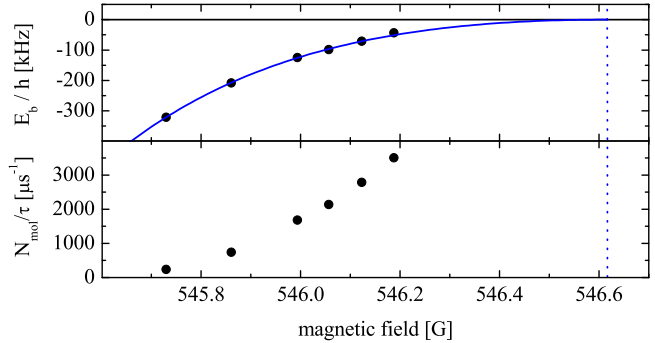}

\caption{The molecular binding energies as a function of magnetic field are shown in the top frame. A value of $B_0 = 546.618(6)\ G$ is obtained for the position of the resonance (dotted line). The bottom shows the obtained number of molecules per pulse length.}

\label{fig2}

\end{figure}

In a first set of experiments, we have studied the efficiency of molecule production for six magnetic fields between $545.73\ G$ and $546.19\ G$. For each magnetic field, the atomic transition frequency of the \K~ atoms was measured as a reference for the association spectra. Figure \ref{fig1} shows the absolute number of KRb molecules versus radio frequency detuning for various magnetic fields. The asymmetry of the association spectra is clearly visible and reflects the relative kinetic energy distribution of the associated atom pairs. The shape of the spectra is very well reproduced by the model described above, where only the binding energy $E_b$ and the scaling factor $\lambda$ were used as fit parameters.

In Figure \ref{fig2}, the extracted binding energies and molecule production rates are plotted as a function of the magnetic field strength. A fit of eq. \eqref{eq:B_of_a} to the data yields $B_0 = 546.618(6)\ G$ (calibration uncertainty $5\ mG$) for the position and $\Delta B = 3.04(2)\ G$ for the width of the resonance. This combination of experimental measurement and theoretical analysis represents the most precise characterization of a heteronuclear resonance in an optical dipole trap, consistent with Ref. \cite{Deuretzbacher2008}. The width of the resonance enters the model through eq. \eqref{eq:chi} and a self-consistent description was obtained by iteration.

Figure \ref{fig4} (a) shows the results for $\lambda$ which represents the total deviation from the theoretical model. It shows that this model quantitatively predicts the heteronuclear molecule yield with good accuracy. The comparably small overall deviation by a factor 2 can be explained as follows: Primarily, collisional relaxation leads to a loss of molecules during the expansion time \cite{Zirbel2008a}. In addition to the overall deviation, the molecule production rate drops for small binding energies close to the Feshbach resonance, since the collisional relaxation is amplified by the resonantly enhanced scattering length. Moreover, the model depends on the temperature, the trapping frequencies and the product of \Rb~ and \K~ atom numbers, thus slight miscalibrations strongly affect the results. 

\begin{figure}[ht!]

\centering

\includegraphics*[width=8.6cm]{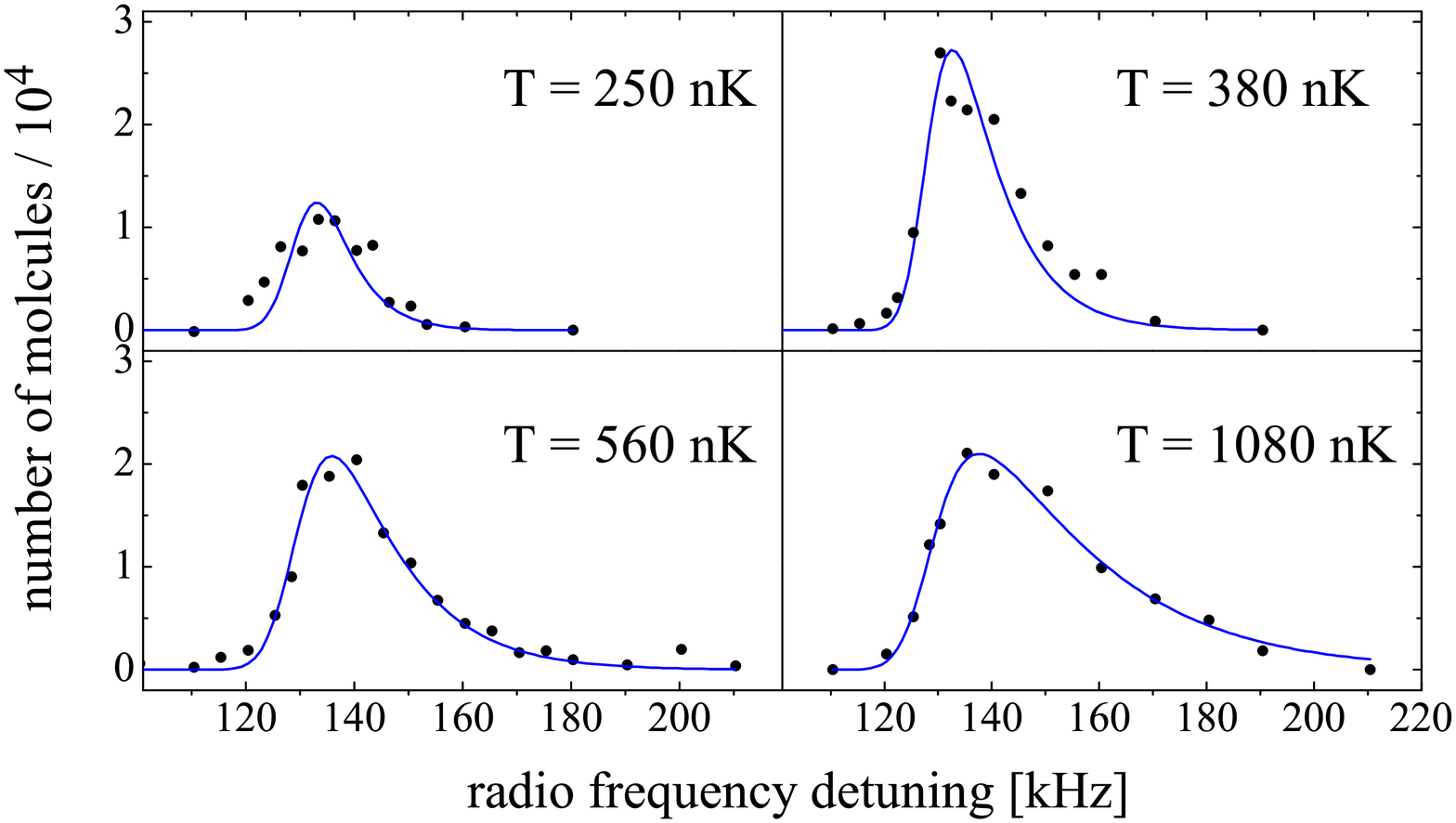}

\caption{Number of associated molecules versus radio frequency for various temperatures. The solid lines show the theoretical model. All measurements were taken at a fixed magnetic field of $545.994\ G$.}

\label{fig3}

\end{figure}

In a second set of experiments, we have investigated the dependence of the molecule production on the temperature of the atomic mixture. By lowering the intensity of the two trapping beams, \K-\Rb-mixtures in a temperature range from $1.1\ \mu K$ down to $250\ nK$ with corresponding mean trapping frequencies of $240$ to $170\ Hz$ ($340$ to $230\ Hz$) for \Rb~ (\K) were produced. The association spectra are shown in Fig. \ref{fig3} with theoretical fits to the data, again using only binding energy $E_b$ and the scaling factor $\lambda$ as free parameters. Once more, the theoretical distribution functions match the experimental findings very well. Since the magnetic field is kept constant at $545.994\ G$, the resulting binding energies of all independent fits yield a value of $127.6(6)\ kHz$, as expected. Figure \ref{fig4} (b) shows the scaling factor $\lambda$ as a function of temperature. It drops slowly with decreasing temperature, since the different quantum statistics of the two species start to play a growing role. While the bosons accumulate in the lowest energy levels, the fermions obey the Pauli principle. This yields a smaller overlap between the clouds and thus a smaller molecule production efficiency. Additionally, the internuclear interaction has been neglected in the occupation densities of the atomic clouds. This repulsive interaction leads to smaller effective trapping frequencies and accordingly to a smaller molecule yield.

\begin{figure}[ht!]

\centering

\includegraphics*[width=8.6cm]{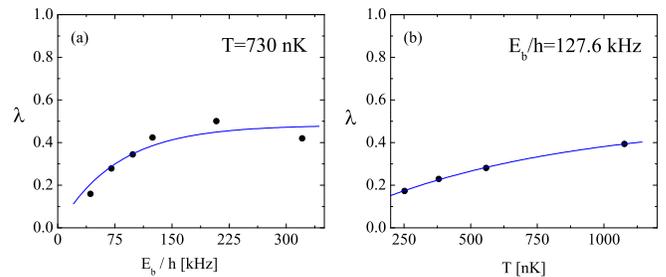}

\caption{Scaling factor versus binding energy (a) and temperature (b). The solid lines are guides the eye.}

\label{fig4}

\end{figure}

In conclusion, we have measured the production efficiency of heteronuclear Feshbach molecules in a wide parameter range of temperatures, binding energies and radio frequencies. A comprehensive analytical model is shown to provide a detailed understanding of the observed spectra and a good quantitative estimate of the absolute production rates. It is of particular relevance for ongoing experiments with heteronuclear Fermi-Fermi mixtures \cite{Taglieber2008,Wille2008}. This work thus paves the way for further experiments using heteronuclear molecules as a starting point for the production of ultracold polar samples.

It will be especially interesting to include the influence of quantum statistics in the model. A numerical approach will be subject of future work. 

We acknowledge support from the Centre for Quantum Engineering and Space-Time Research QUEST and from the Deutsche Forschungsgemeinschaft (SFB 407, European Graduate College Interference and Quantum Applications).

Note added: Recently, we have become aware of a complementary work with a bosonic heteronuclear mixture of $^{41}K$~and $^{87}Rb$ \cite{Weber2008}.

\bibliographystyle{prsty}
\bibliography{moleculepaper}
\end{document}